\documentclass[%
 reprint,
superscriptaddress,
nobibnotes,
 amsmath,amssymb,
 aps,
prb,
showkeys
]{revtex4-2}
\usepackage{graphicx}  
\usepackage{dcolumn}   
\usepackage{bm}        
\usepackage{amssymb}   
\usepackage{siunitx}
\usepackage[breaklinks=true,colorlinks,citecolor=blue,linkcolor=red,urlcolor=blue]{hyperref}
\usepackage{multirow}
\usepackage{amsmath}
\usepackage{hhline}
\usepackage{lipsum}

\usepackage{soul}
\usepackage{color} 

\usepackage{braket}
\usepackage{autobreak}

\usepackage{amsfonts}
\usepackage{amsmath}
\usepackage{slashed}
\usepackage{amsthm}
\usepackage{amssymb}

\usepackage[dvipsnames]{xcolor}

\definecolor{newcolor}{rgb}{0.0, 0.6, 0.0}
\definecolor{mycolor}{rgb}{0.0, 0.0, 0.8}
\definecolor{red}{rgb}{1, 0.0, 0.0}
\definecolor{blue}{rgb}{0.0, 0.0, 1.0}

\begin{document}

\title{Tunable Competing Electronic Orders in Double Quantum Spin Hall Superlattices}

\author{Yi-Chun Hung}
\email[Contact author: ]{hung.yi@northeastern.edu}
\affiliation{Department of Physics, Northeastern University, Boston, Massachusetts 02115, USA}
\affiliation{Quantum Materials and Sensing Institute, Northeastern University, Burlington, Massachusetts 01803, USA}

\author{Chen-Hsuan Hsu}
\affiliation{Institute of Physics, Academia Sinica, Taipei 115201, Taiwan}

\author{Arun Bansil}
\email[Contact author: ]{ar.bansil@northeastern.edu}
\affiliation{Department of Physics, Northeastern University, Boston, Massachusetts 02115, USA}
\affiliation{Quantum Materials and Sensing Institute, Northeastern University, Burlington, Massachusetts 01803, USA}


\begin{abstract}
Competing superconducting (SC) and density-wave orders are of key importance in generating unconventional superconductivity and emergent electronic responses. Quasi-one-dimensional models provide insight into these competing orders and suggest higher-dimensional realizations through coupled-wire constructions, but analysis of such systems remains limited. Recent studies suggest that double helical edge states (DHESs) in double quantum spin Hall insulators (DQSHIs) form a two-channel Luttinger liquid that exhibits SC and spin density wave (SDW) phases and their $\pi$-junction analogs. Here, we analyze weakly coupled DHESs from the surface of a periodically stacked layered structure consisting of DQSHIs and dielectrics, where inter-edge interactions approximately develop a tunable helical sliding Luttinger liquid (HSLL) order. Using a renormalization-group analysis, we construct phase diagrams and identify a regime of HSLL parameters that favor competing two-dimensional $\pi$-SC and $\pi$-SDW orders. We identify parameter regimes where the competing orders could be realized experimentally in nanoscale devices. Our study suggests a promising materials platform for exploring tunable $\pi$-SC and $\pi$-SDW orders in double quantum spin Hall superlattices.
\end{abstract}

\maketitle

\section{Introduction}
\par Strongly correlated electronic systems are usually characterized by complex phase diagrams with multiple competing orders. Examples include high-temperature superconductors \cite{Chang2012, RevModPhys.87.457} and kagome materials which exhibit complex interplay of superconducting (SC), charge density wave (CDW) and other orders \cite{Kang2023, Yu2021, Wilson2024, Song2023, PhysRevResearch.3.043018, PhysRevLett.127.237001, PhysRevLett.127.046401, PhysRevResearch.5.L012032}. The coexistence or competition of SC and spin density wave (SDW) phases is another hallmark of unconventional superconductivity in iron-based compounds \cite{Cai2013, PhysRevX.3.011020, kordyuk2012}, nickelates \cite{Khasanov2025, Lane2023}, organics \cite{PhysRevB.30.1570, Vuletic2002}, and magic-angle twisted bilayer graphene \cite{doi:10.1126/science.abc2836}. Understanding the interplay between SC and SDW, however, has remained challenging due to the limited tunability of existing materials platforms.

\par In this connection, quasi-one-dimensional (quasi-1D) models have been extensively studied as they naturally exhibit non-Fermi liquid physics. Various theoretical proposals have explored the emergence of SC and SDW phases and the conditions under which they coexist or compete \cite{PhysRevB.91.195102, PhysRevB.79.224520, Duprat2001, Palistrant1985}. Higher-dimensional correlated systems can be obtained by coupling multiple 1D wires or chains \cite{book:Giamarchi, xie2025cdwsc, PhysRevB.74.064405, PhysRevB.108.245138} to offer greater control through variation of intra- and inter-wire interactions. Organic superconductors \cite{PhysRevB.30.1570, Vuletic2002}  present a promising example of coupled-wire systems \cite{ PhysRevB.61.16393}  for coexisting SC and SDW phases. 

\par Recent studies of double helical edge states (DHESs) in double quantum spin Hall insulators (DQSHIs) indicate that these states can host a higher-order topological phase \cite{PhysRevB.108.245103, PhysRevB.104.L201110, PhysRevB.110.035161, PhysRevB.110.035125}, and may also stabilize new phases supporting Majorana Kramers pairs \cite{PhysRevB.111.245145}. Originating from the non-trivial band topologies found in DQSHIs \cite{PhysRevB.80.125327, shulman2010robust, wang2023feature, Lin2024, PhysRevLett.106.106802}, the associated DHESs have been identified in numerous materials for potential realization \cite{PhysRevB.109.155143, lefeuvre2025, Wang_2024, PhysRevB.110.L161104, PhysRevB.110.035161, PhysRevB.94.235111, Kang:2024a, Kang:2024b, PhysRevB.110.195142, Dziawa2012, Tanaka2012, Liu2014, Liu2015}. Unlike the single helical edge states, interactions transform DHESs into a two-channel Luttinger liquid (LL) \cite{PhysRevB.111.245145}, which resembles the Luther-Emery liquid characterized by a gapless charge sector and a gapped spin sector \cite{PhysRevLett.33.589, book:Giamarchi}. Collective excitations within the DHESs remain helical due to the helical nature of the underlying single-particle excitations. Considering the potential of coupled Luther-Emery liquid arrays for supporting competing two-dimensional (2D) SC and CDW phases \cite{book:Giamarchi, xie2025cdwsc}, it is natural to ask if an array of coupled DHESs could also support such competing phases.

\begin{figure}[t]
  \centering
  \centering
    \includegraphics[width=\linewidth]{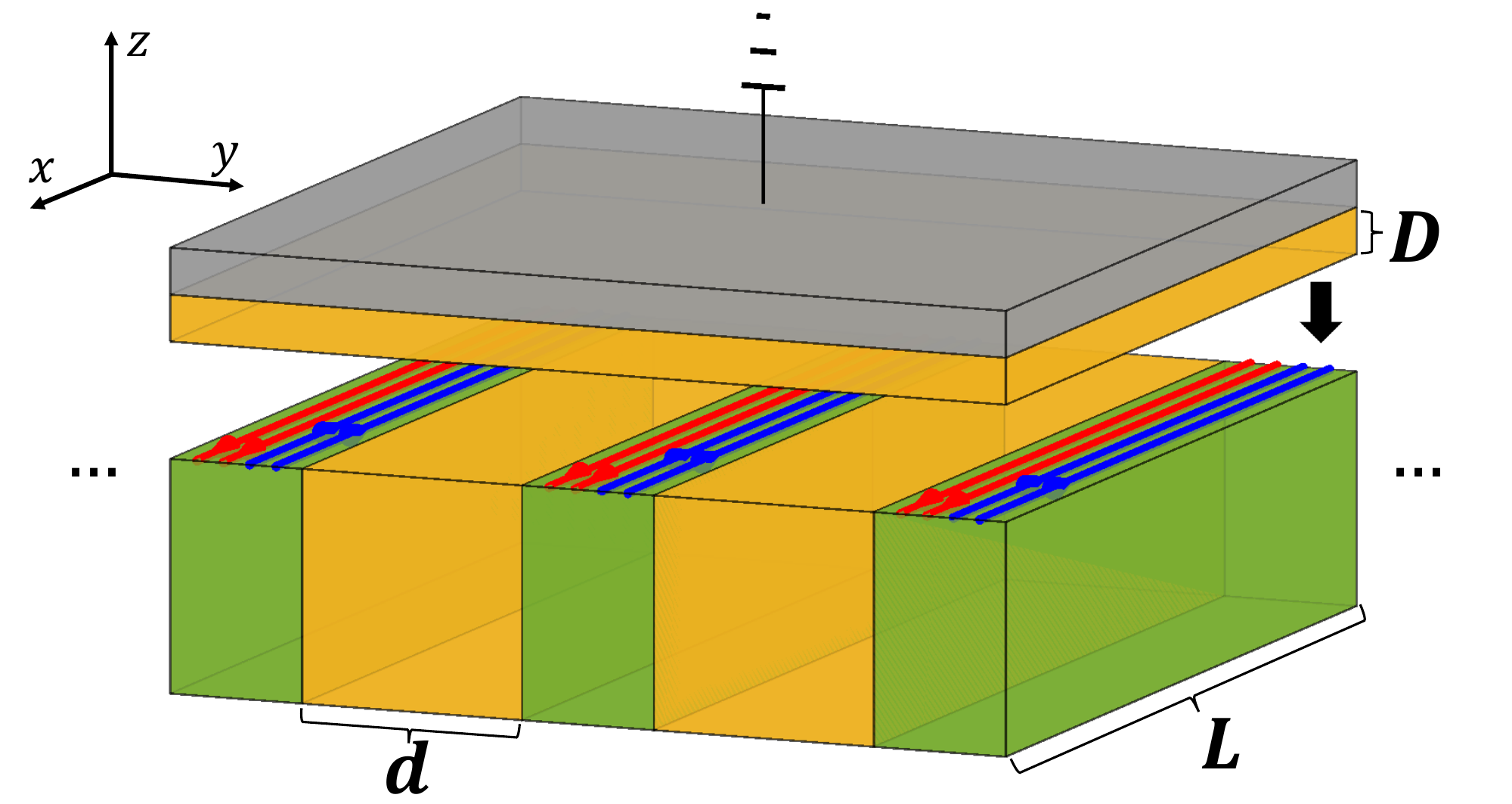}
  \caption{A schematic sectional view of the double quantum spin Hall superlattice. The grey, green, and yellow regions indicate the grounded metallic gate, the DQSHIs, and the embedding dielectric materials, respectively. Red and blue arrows indicate the up- and down-spin states forming the DHESs. Black arrow indicates the direction of coverage, highlighting the interface between the two parts.
  }
  \label{fig:01}
\end{figure}

\par In this study, we analyze weakly coupled DHESs emerging from the surface of a double quantum spin Hall superlattice of length $L$, created by embedding dielectric materials of thickness $d$ periodically between DQSHIs along the out-of-plane direction. Next, the superlattice is coated with a dielectric material of thickness $D$, and a grounded metallic gate is added on top, see Fig.~\ref{fig:01}. We show that the inter-edge interaction approximately develops a helical sliding Luttinger liquid (HSLL) order with parameters tunable by the embedded dielectric materials. Importantly, we demonstrate that symmetry-allowed inter-edge tunnelings suggest competing 2D $\pi$-SC and $\pi$-SDW phases over a certain range of parameters through a renormalization group (RG) analysis to the lowest order. We discuss the potential materials realization of DQSHIs. Our system offers a versatile platform for realizing competing $\pi$-SC and $\pi$-SDW phases.

\section{Setup}
\subsection{The Double Helical Edge States (DHESs)}\label{sec:1A}
\par
\begin{figure}[t]
  \centering
    \includegraphics[width=\linewidth]{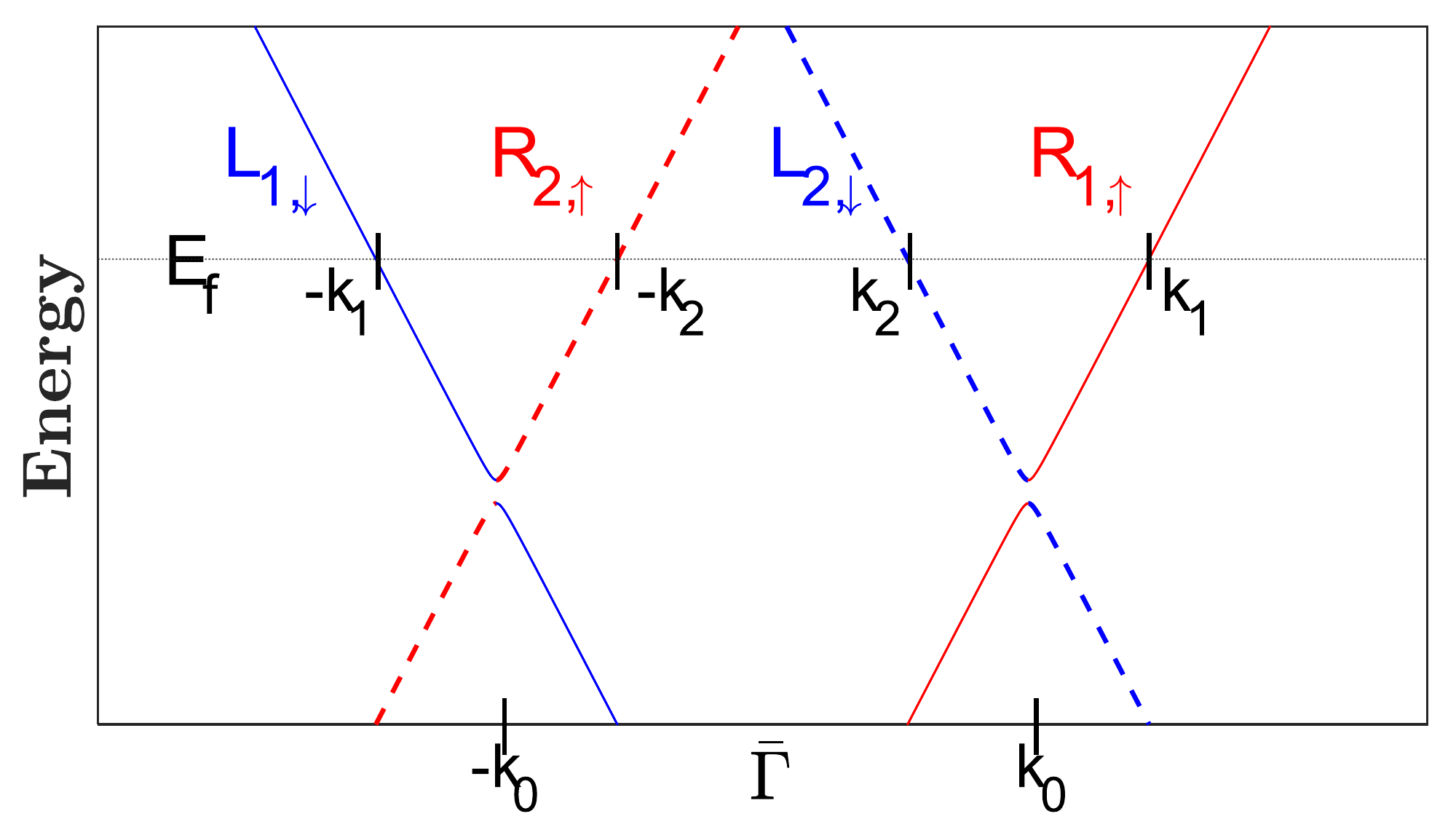}
  \caption{A schematic of the dispersion of DHESs near the Fermi level, $\text{E}_\text{f}$. The right-(left-)moving modes are labeled as $R_{\mu,\uparrow}$ ($L_{\mu,\downarrow}$), where the index $\mu=1,2$ denotes the two time-reversal sectors represented by the solid and dashed bands. $k_0\equiv (k_1+k_2)/2$ denotes the midpoint of the edge states' momentum. 
  }
  \label{fig:02}
\end{figure}

 DHESs host two pairs of helical edge states, which transform independently under time-reversal symmetry \cite{PhysRevB.110.035125}. These low-energy states can be represented by the fermionic fields:
\begin{equation}\label{eq:00}
    \psi_\mu(r) = R_{\mu\uparrow}(r)e^{-i(-1)^{\mu} k_\mu r} + L_{\mu\downarrow}(r)e^{i(-1)^{\mu} k_\mu r},
\end{equation}
where the index $\mu=1,2$ indicates the time-reversal sectors (Fig.~\ref{fig:02}). We suppress the spin index in Eq.~\eqref{eq:00} from now on for simplicity.

\par When the forward scatterings are included in the DHESs, the system develops a two-channel LL featuring a gapless charge sector and a gapped pseudospin sector. Specifically, a mass term, arising from $e$-$e$ interactions or spin-orbit coupling, induces the gap in the pseudospin sector with the following bosonization representation:
\begin{align}\label{eq:01}
    H_{m} = \frac{m}{2(\pi a)^2}\int dr\cos[2\sqrt{2}\vartheta_\tau(r)],
\end{align}
where the index $\tau$ indicates the pseudospin sector; see Appendix \ref{appx:A} for details of the bosonization convention. In contrast to the single helical edge states~\footnote{See, e.g., Ref.~\cite{Hsu_2021} for a review of various relevant backscattering mechanisms.}\nocite{Hsu_2021}, such a mass term does not require a broken time-reversal symmetry in DHESs and originates from the forward scattering between different time-reversal sectors tied to the long-range component of the interacting potential \cite{PhysRevB.111.245145} or the second-order effect from the spin-orbit coupling; see Appendix \ref{appx:A2} for details.

\par
\begin{figure}[t]
  \centering
  \centering
    \includegraphics[width=\linewidth]{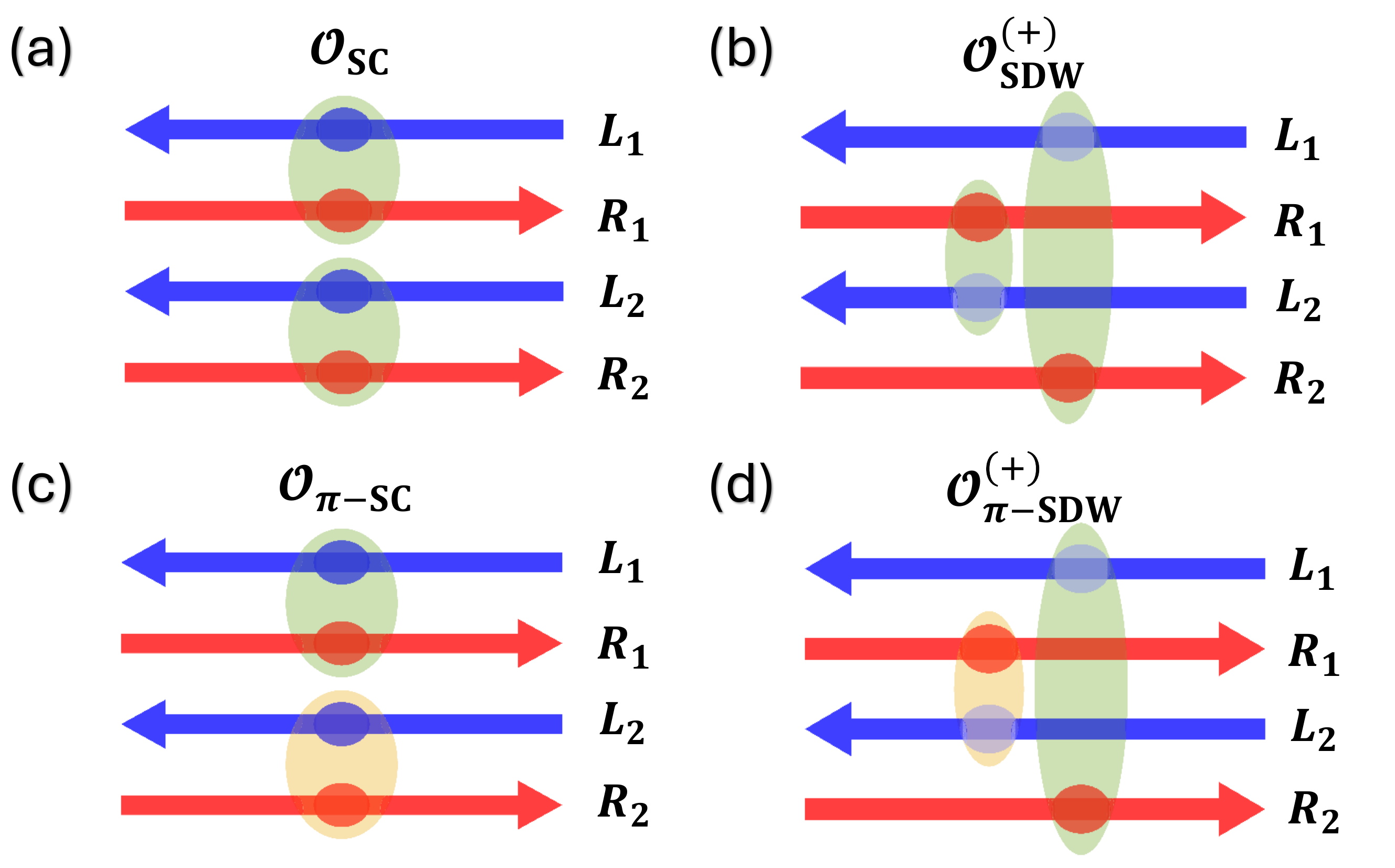}
  \caption{Schematic diagrams illustrating excitations in (a) SC, (b) SDW, (c) $\pi$-SC, and (d) $\pi$-SDW phases. Only the excitations corresponding to the $\mathcal{O}_{(\pi\text{-})\text{SDW}}^{(+)}$ for SDW phases are shown without loss of generality. Filled (empty) spheres represent electrons (holes). Green and yellow clouds around the two spheres indicate the formation of electron-electron pairs for ($\pi$-)SC or electron-hole pairs for ($\pi$-)SDW phases. There is a $\pi$-phase difference between the pairs with green and yellow clouds corresponding to Eqs.~\eqref{eq:03} and \eqref{eq:04}.
  }
  \label{fig:03}
\end{figure}

 In the presence of the mass term, $2\sqrt{2}\vartheta_\tau(r)\mod2\pi$ becomes pinned at $0,\pi$ on the mesoscopic scale and gives rise to phases like SC and SDW for $m<0$, and their $\pi$-junction counterparts ($\pi$-SC and $\pi$-SDW) for $m>0$; the nomenclature denotes a $\pi$-phase difference between time-reversal sectors. The operators characterizing the corresponding instabilities are:
\begin{align}
\mathcal{O}_{(\pi\text{-})\text{SC}}(r) & \propto R_{1}^\dagger(r) L_{1}^\dagger(r) - L_{1}^\dagger(r) R_{1}^\dagger(r) \pm (1\leftrightarrow2) , \label{eq:03}
\\ \mathcal{O}_{(\pi\text{-})\text{SDW}}^{(+)}(r) & \propto  R_{2}^\dagger(r)L_{1}(r)  \pm (1\leftrightarrow2) , \label{eq:04}
\\ \mathcal{O}_{(\pi\text{-})\text{SDW}}^{(-)}(r) & \propto  L_{1}^\dagger(r) R_{2}(r)  \pm (1\leftrightarrow2), \label{eq:05}
\end{align}
where $\mathcal{O}_{(\pi\text{-})\text{SDW}}^{(\pm)}\equiv\mathcal{O}_{(\pi\text{-})\text{SDW}}^{(x)}\pm i\mathcal{O}_{(\pi\text{-})\text{SDW}}^{(y)}$ characterize the instabilities of in-plane components of ($\pi$-)SDW. A schematic diagram of the related excitations is shown in Fig.~\ref{fig:03}. Note that the four schematics in Fig.~\ref{fig:03}, along with their Hermitian conjugates, represent the relevant scattering processes between the left- and right-moving modes in the DHESs within a single edge channel. Other processes do not correspond to RG-relevant instabilities with divergent susceptibilities \cite{PhysRevB.111.245145}.

\subsection{Inter-edge Tunnelings}
\par Upon coupling DHESs on different edges in an array (Fig.~\ref{fig:01}), the single-particle inter-edge tunneling becomes irrelevant due to the gapped pseudospin sector \cite{book:Giamarchi}. The remaining relevant tunnelings are essentially two-body inter-edge tunnelings in the SC and SDW channels. Due to time-reversal symmetry, the allowed two-body inter-edge tunnelings are of the form:
\begin{align}
H_{(\pi\text{-})\text{SC}}^{(\text{inter})} & = -\sum_{i\neq j}J_{\text{SC},|i-j|} \notag
\\ & \quad \quad  \times \int dr\mathcal{O}^\dagger_{(\pi\text{-})\text{SC},i}(r)\mathcal{O}_{(\pi\text{-})\text{SC},j}(r) ,
\\ H_{(\pi\text{-})\text{SDW}}^{(\text{inter})} & = -\sum_{i\neq j}J_{\text{SDW},|i-j|} \notag
\\ & \quad \quad \times \int dr \mathcal{O}_{(\pi\text{-})\text{SDW},i}^{(+)}(r)\mathcal{O}_{(\pi\text{-})\text{SDW},j}^{(-)}(r).
\end{align}
Here, $J_{\text{SC},|i-j|}$ and $J_{\text{SDW},|i-j|}$ indicate the strength of the inter-edge tunneling for the ($\pi$-)SC and ($\pi$-)SDW channels between $i$th and $j$th edges, and $\mathcal{O}_{(\pi\text{-})\text{SC},i}$ and $\mathcal{O}_{(\pi\text{-})\text{SDW},i}^{(\pm)}$ denote operators characterizing instability of ($\pi$-)SC and the in-plane components of ($\pi$-)SDW on the $i$th edge. The ($\pi$-)SC and ($\pi$-)SDW channels represent pair tunneling with particle-particle and particle-hole pairs in the charge sector, respectively \cite{book:Giamarchi, PhysRevLett.86.676}. Due to the helical nature of the charge sector, the particle-hole pairs resemble the electronic spin \cite{PhysRevB.111.245145}. Thus, the corresponding inter-edge tunneling develops a 2D ($\pi$-)SDW phase when it becomes  relevant. 

\par The inter-edge tunnelings have the following bosonization representations:
\begin{align}
H_{(\pi\text{-})\text{SC}}^{(\text{inter})} & \sim -\sum_{i\neq j}\frac{J_{\text{SC},|i-j|}}{(\pi a)^2}\int dr e^{i\sqrt{2}\vartheta_{c,i}(r)}e^{-i\sqrt{2}\vartheta_{c,j}(r)} \notag
\\ & \quad \quad \quad \times f_{(\pi)}[\sqrt{2}\vartheta_{\tau,i}(r)]f_{(\pi)}[\sqrt{2}\vartheta_{\tau,j}(r)] ,
\\ H_{(\pi\text{-})\text{SDW}}^{(\text{inter})} & \sim -\sum_{i\neq j}\frac{J_{\text{SDW},|i-j|}}{(\pi a)^2}\int dr e^{i\sqrt{2}\varphi_{c,i}(r)}e^{-i\sqrt{2}\varphi_{c,j}(r)} \notag
\\ & \quad \quad \quad \times f_{(\pi)}[\sqrt{2}\vartheta_{\tau,i}(r)]f_{(\pi)}[\sqrt{2}\vartheta_{\tau,j}(r)],
\end{align}
where the index $c$ indicates the charge sector and $f(x)=\cos(x)$ with $f_{\pi}(x)=\sin(x)$. Note that the gap in the pseudospin sector on each edge is developed by Eq.~\eqref{eq:01}. Therefore, for $m>0$, $\sqrt{2}\vartheta_\tau(r)$ is pinned at $(n+1/2)\pi$, $H_{\pi\text{-SC}}^{(\text{inter})}$ and $H_{\pi\text{-SDW}}^{(\text{inter})}$ are favored. Similarly, $H_{\text{SC}}^{(\text{inter})}$ and $H_{\text{SDW}}^{(\text{inter})}$ are favored for $m<0$. Deep in the phase where $m$ is relevant, we can absorb the fields in the pseudospin sector by their mesoscopic expectation values \cite{book:Giamarchi} and further simplify the inter-edge tunnelings into:
\begin{align}
H_{(\pi\text{-})\text{SC}}^{(\text{inter})} & = -\sum_{i\neq j}\frac{J_{\text{SC},|i-j|}}{(\pi a)^2}\int dr e^{i\sqrt{2}[\vartheta_{c,i}(r)-\vartheta_{c,j}(r)]} ,  \label{eq:H_int_SC}
\\ H_{(\pi\text{-})\text{SDW}}^{(\text{inter})} & = -\sum_{i\neq j}\frac{J_{\text{SDW},|i-j|}}{(\pi a)^2}\int dr  e^{i\sqrt{2}[\varphi_{c,i}(r)-\varphi_{c,j}(r)]}. \label{eq:H_int_SDW}
\end{align}

\subsection{Helical Sliding Luttinger Liquid}
\par Besides the inter-edge tunneling, different edges are also coupled through inter-edge forward scattering. Note that screened Coulomb interactions modify only the charge-sector interaction parameters. Upon a Fourier transform in the transverse direction, we obtain a helical sliding Luttinger liquid (HSLL) described by, 
\begin{align}\label{eq:HSLL}
    H_{\text{HSLL}} = & \int dr\int_{-\pi}^\pi \frac{dk_\perp}{2\pi} \,\frac{\hbar u}{2\pi} [K^{-1}_{k_\perp}(\partial_r\varphi_{k_\perp}(r))^2 \notag \\ & \quad \quad \quad \quad \quad \quad \quad \quad + K_{k_\perp}(\partial_r\vartheta_{k_\perp}(r))^2],
\end{align}
which can be viewed as a helical version of the sliding Luttinger liquid \cite{PhysRevLett.86.676} or a smectic metal \cite{PhysRevLett.85.2160}.
Here $K_{k_\perp}$ denotes the HSLL parameter for the $\varphi_{k_\perp}(r)$ and $ \vartheta_{k_\perp}(r)$ fields with dimensionless transverse momentum $k_\perp$, and $u$ denotes the renormalized velocity. 

Considering a heterostructure with sufficiently weak inter-edge couplings and a sufficiently large inter-edge distance, we expand the HSLL parameter $K_{k_\perp}$ to the first harmonic of the transverse momentum; see Appendix \ref{appx:B} for details:
\begin{equation}
    K_{k_\perp} \cong K_0 \big[ 1 + K_1\sin(|k_\perp|) \big],
\end{equation}
where
\begin{align}
K_{0} & = \sqrt{\frac{1}{1 + \frac{2e^2D}{\hbar v\pi\epsilon d}}}, \label{eq:K0}
\\ 
K_{1} & = \frac{2e^2D^2}{\hbar v\pi\epsilon d^2 }. \label{eq:K1}
\end{align}
Here, $v$ is the Fermi velocity of the helical edge states, $\epsilon$ refers to the dielectric constant of the double quantum spin Hall superlattice, $d$ is the inter-edge distance, and $D$ is the distance between the edge states and the metallic gate. A large inter-edge distance ($d\gg D$) indicates a well-screened system. We assume similar dielectric constants for the DQSHI and the dielectric material in Eqs.~\eqref{eq:K0} and \eqref{eq:K1} for simplicity. 

\par Given a sufficiently large inter-edge distance, only the nearest-neighbor inter-edge tunnelings, $J_{\text{SC},1}$ and $J_{\text{SDW},1}$ are relevant, although it is straightforward to include terms beyond the nearest-neighboring DQSHI edges. We show below how inter-edge tunnelings lead to competing 2D ($\pi$-)SC and ($\pi$-)SDW phases.

\section{Phase diagram} 
\par
\begin{figure}[t]
  \centering
  \centering
    \includegraphics[width=\linewidth]{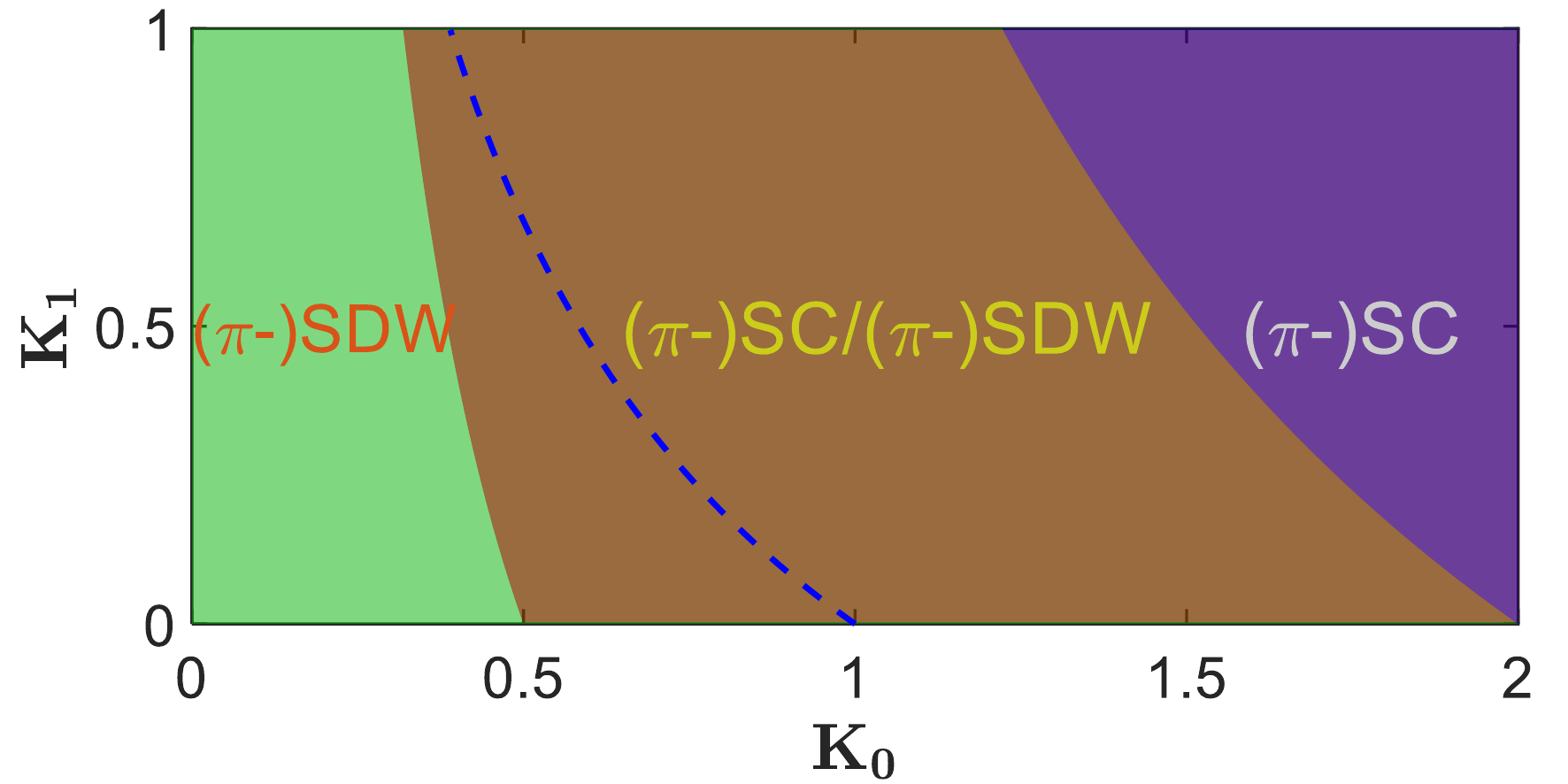}
  \caption{Generic phase diagram as a function of the interaction parameters: green, brown, and purple colors indicate the regions of ($\pi$-)SDW, competing ($\pi$-)SDW and ($\pi$-)SC, and ($\pi$-)SC phases, respectively. Blue dashed line shows where the scaling dimensions of operators for ($\pi$-)SDW and ($\pi$-)SC phases are the same. 
  }
  \label{fig:04}
\end{figure}

 We derive the RG flow equations for the two-body inter-edge tunneling from co-tunneling processes up to the leading order \cite{book:Giamarchi, PhysRevB.97.045415}: 
\begin{align}
    \frac{d\tilde{J}_{\text{SC},1}}{dl} & = \bigg[ 2-\frac{1}{K_0}\frac{2\cos^{-1}(K_1)}{\pi\sqrt{1-K_1^2}} \bigg]\tilde{J}_{\text{SC},1} , \label{eq:11}
    \\ \frac{d\tilde{J}_{\text{SDW},1}}{dl} & = \bigg[ 2- K_{0}\Big( 1 + \frac{2K_1}{\pi} \Big) \bigg]\tilde{J}_{\text{SDW},1}, \label{eq:12}
\end{align}
where $l\equiv\ln [ a(l)/a(0) ]$, $a$ is the cutoff of the low-energy theory, and $\tilde{J}_{[...],N}\equiv J_{[...],N}/\hbar u\pi^2$ are dimensionless two-body inter-edge tunnelings; see Appendix \ref{appx:C} for details. According to Eqs.~\eqref{eq:11} and \eqref{eq:12}, the $\tilde{J}_{(\pi\text{-})\text{SC},1}$ is RG relevant for $K_0>\cos^{-1}(K_1)/\big(\pi\sqrt{1-K_1^2}\big)$. The $\tilde{J}_{(\pi\text{-})\text{SDW},1}$ is RG relevant for $K_0<2/(1+2K_1/\pi)$. 
This implies a regime with $\cos^{-1}(K_1)/\big(\pi\sqrt{1-K_1^2}\big)<K_0<2/(1+2K_1/\pi)$, in which both the ($\pi$-)SC and ($\pi$-)SDW phases are RG relevant and compete. Figure~\ref{fig:04} summarizes the resulting phase diagram. The determination of whether the competing SC and SDW phases or the $\pi$-SC and $\pi$-SDW phases are at play depends on the sign of the mass in Eq.~\eqref{eq:01}. We emphasize that although higher-order RG analysis can in principle provide more precise phase boundaries, it is not expected to introduce new phases or new regimes of competing orders.

\par Beyond the phase diagram, a comprehensive RG analysis also uncovers the source of two-body inter-edge tunnelings, which can arise from the second-order contributions of single-particle inter-edge tunneling $t_\perp$ (co-tunneling). This is clearly reflected in the schematic RG flow equations $d\tilde{J}/dl=(2-\eta)\tilde{J}+\mathcal{O}(t_\perp^2)$, see Appendix \ref{appx:C} for details. In this way, even though $\tilde{J}_{(\pi\text{-})\text{SC},1}(l=0)$ and $\tilde{J}_{(\pi\text{-})\text{SDW},1}(l=0)$ are initially zero, non-zero values of $\tilde{J}_{(\pi\text{-})\text{SC},1}$ and $\tilde{J}_{(\pi\text{-})\text{SDW},1}$ can emerge through the RG flow due to $t_{\perp}(l=0) \neq 0$, although $t_\perp$ is considered RG irrelevant due to the gapped pseudospin sector \cite{book:Giamarchi}.
  
\section{Potential platforms}
\par We now turn to discuss how we could realize heterostructures that host HSLL and competing SC and SDW orders. We first consider the DHESs whose existence has been experimentally and theoretically demonstrated in various topological materials such as spin and mirror Chern insulators with the associated spin and mirror Chern numbers of $\pm2$. Experimentally realized platforms include bilayer $\beta$-Bi$_4$Br$_4$ \cite{PhysRevB.109.155143, lefeuvre2025}, twisted bilayer MoTe$_2$ \cite{Kang:2024a, mao2025} and WSe$_2$ \cite{Kang:2024b}, thin films of the SnTe family \cite{Dziawa2012, Tanaka2012, Liu2014, Liu2015}, and coupled HgTe quantum wells \cite{PhysRevB.85.125309, PhysRevB.101.241302, PhysRevB.103.L201115, PhysRevB.93.235436, Ferreira2022, Krishtopenko2016}. Additional candidate materials proposed by first-principles calculations include $\alpha$-antimonene \cite{Wang_2024}, RuBr$_3$ \cite{PhysRevB.110.L161104, PhysRevB.110.035161}, the oxide compound $(\text{M}\text{O}_2)_2(\text{Zr}\text{O}_2)_4$ with M = Pt, W \cite{PhysRevB.94.235111}, and Na$_2$CdSn \cite{PhysRevB.110.195142}. Notably, the spin-valley locking in transition-metal dichalcogenides (TMDs) such as MoTe$_2$ and WSe$_{2}$ ensures spin-$U(1)$ symmetry in their twisted bilayers, facilitating the realization of the DHESs.

\par
\begin{figure}[h]
  \centering
  \centering
    \includegraphics[width=\linewidth]{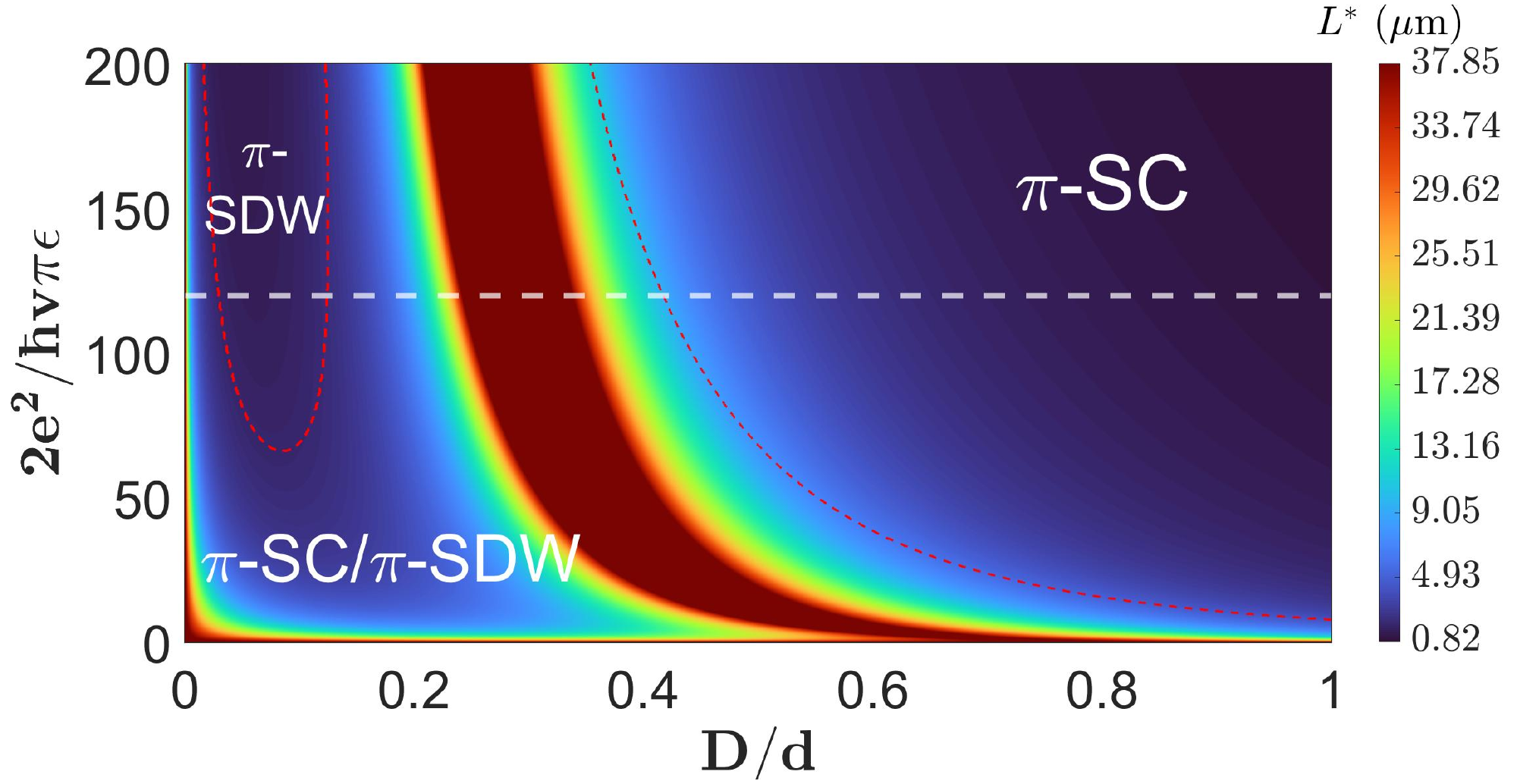}
  \caption{
Phase diagram and minimal edge length $L^*$ in Eq.~\eqref{eq:L} estimated with $a(0)\approx4$ nm and $\tilde{J}_{\alpha}(l\to0)\approx10^{-4}$ as functions of system parameters defined in Eqs. \eqref{eq:K0} and \eqref{eq:K1}. Red dashed lines mark phase boundaries, and the colormap indicates the $L^*$ values below which the system remains in the HSLL phase. The white dashed line marks an estimated (order of magnitude) parameter regime relevant for twisted-bilayer-TMD based double quantum spin Hall superlattices using MoTe$_2$ and WSe$_2$.
  }
  \label{fig:06}
\end{figure}

\begin{figure}[h]
  \centering
  \centering
    \includegraphics[width=\linewidth]{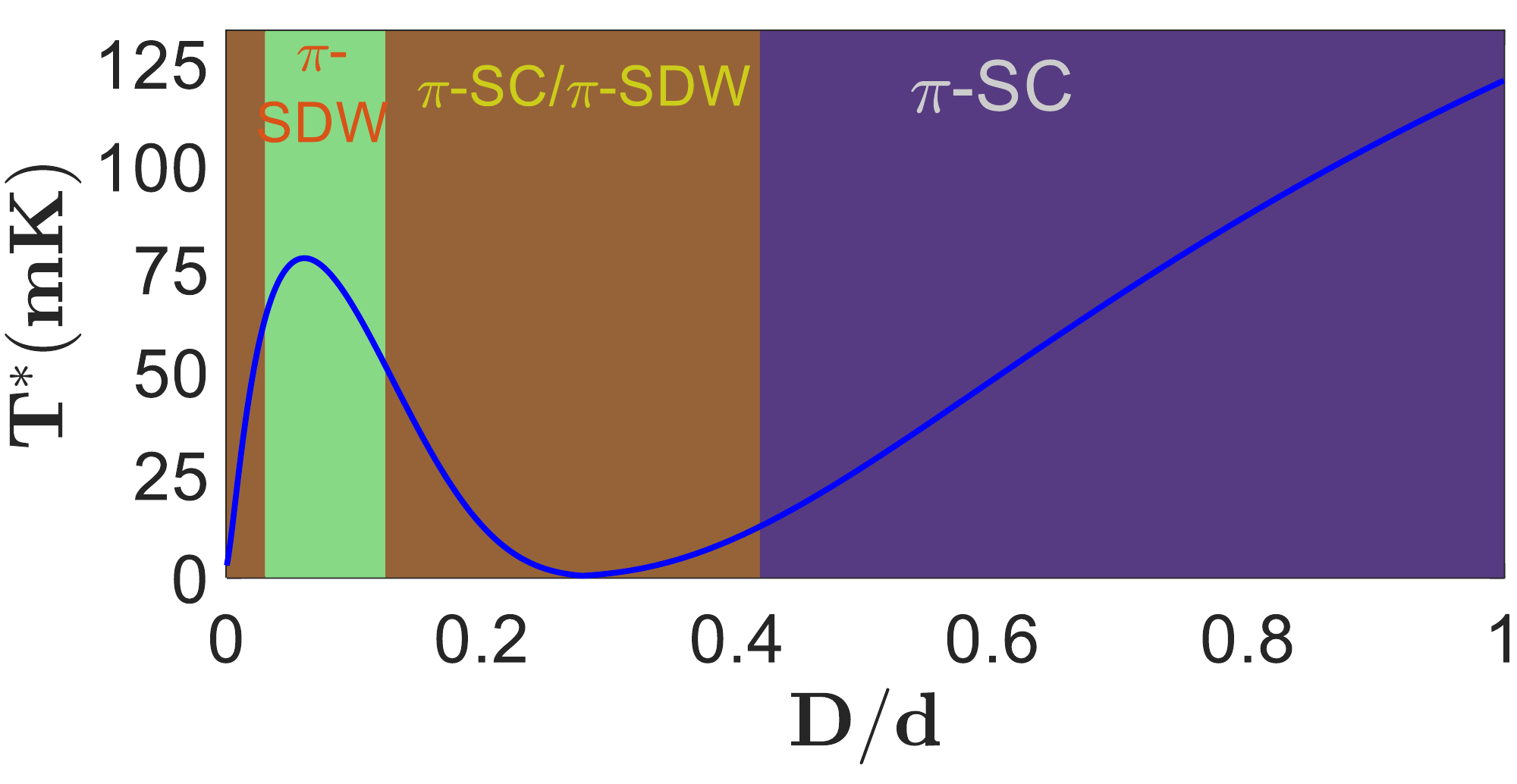}
  \caption{Characteristic temperature $T^*$ for various phases in the twisted-bilayer MoTe$_2$- or WSe$_2$-based double quantum spin Hall superlattices, estimated from the parameter set corresponding to the white dashed line in Fig.~\ref{fig:06}.
 Colored regions correspond to the phases indicated in Fig.~\ref{fig:04}.
  }
  \label{fig:07}
\end{figure}

 Note that the predominant screened Coulomb interaction would lead to $m>0$ in the pseudospin sector \cite{PhysRevB.111.245145} and thus favor the $\pi$-SC and $\pi$-SDW phases, as discussed in Sec.~\ref{sec:1A}. But, for a given DQSHI, the HSLL parameters given in Eqs.~\eqref{eq:K0} and \eqref{eq:K1} could be straightforwardly adjusted through the inter-edge separation $d$ and the screening distance to the metallic gate $D$ such that $1/\pi< K_0<1$ and $0<K_1\ll 1$, driving the system into the regime of phase competition. This expectation can be further substantiated by estimating the interaction strength based on material-specific parameters for MoTe$_2$ and WSe$_{2}$ \cite{mao2025, Bi2021, Laturia2018}; the corresponding phase diagram is shown in Fig.~\ref{fig:06}. Assuming $D/d \ll 1$ in Eqs.\eqref{eq:K0} and \eqref{eq:K1}, and using $\hbar v \approx \mathcal{O}(10^{-11})\text{ eV}\cdot\text{m}$ \cite{mao2025} and $\epsilon \approx 10\epsilon_0$ \cite{Bi2021, Laturia2018} for twisted bilayer MoTe$_2$ and WSe$_2$, our analysis reveals the presence of competing $\pi$-SC and $\pi$-SDW phases over a wide parameter range, see Fig.~\ref{fig:06}. Based on the RG flow in Eqs.~\eqref{eq:11} and \eqref{eq:12}, we can determine the minimal edge length $L^*$ necessary to reach the strong-coupling regime, where the dimensionless inter-edge tunnelings attain $\mathcal{O}(1)$:
\begin{equation}\label{eq:L}
    L^* = \min_{\alpha\in\{\text{SC},\text{SDW}\}}\bigg( a(0)[\tilde{J}_{\alpha}(l\to0)]^{-\frac{1}{2-\eta_{\alpha}}} \bigg),
\end{equation}
where $\eta_{\text{SC}}=2\cos^{-1}(K_1)/\pi K_0\sqrt{1-K_1^2}$ and $\eta_{\text{SDW}}=K_{0}\big( 1 + 2K_1/\pi)$. To estimate $L^*$ in twisted-bilayer-TMD-based double quantum spin Hall materials, we use $a(0)\approx4$ nm \cite{mao2025} and $\tilde{J}_{\alpha}(l\to0)\approx10^{-4}$. This choice follows from estimating $\tilde{J}_{\alpha}(l\to0)$ as $ (t_\perp(l=0)a/\hbar v)^2$, with $t_\perp\approx10^{-2}\hbar v/a$ chosen to align with the energy scale hierarchy in Eq.~\eqref{eq:20}, see Appendix \ref{appx:C} and the following section for details. The associated characteristic temperature $T^*$ can also be determined by $T^*=\hbar v/k_B L^*$, marking the temperature scale below which the system enters the strong-coupling regime. Figures~\ref{fig:06} and \ref{fig:07} summarize the preceding estimates and demonstrate that a rich phase diagram can arise from competing $\pi$-SC and $\pi$-SDW phases below $O(100~\textrm{mK})$ in devices larger than $O(\mu\textrm{m})$.

\par Regarding the energy hierarchy in our analysis, note that we have explored weak inter-edge couplings, focusing solely on the nearest-neighbor tunnelings, and expanding the HSLL parameter to the first harmonic of the transverse momentum. Weak inter-edge couplings ensure that the gap in the pseudospin sector develops first to avoid a dimensional crossover, yielding the energy scale hierarchy:
\begin{equation}\label{eq:20}
    \frac{\hbar v}{a}\gg U_{2k_0}\gg t_\perp.
\end{equation} 
Here, $v$ and $a$ are the Fermi velocity and cutoff of the low-energy theory, respectively. $U_q$ is the Fourier component of the intra-edge density-density interaction with momentum $q$. The hierarchy in Eq.~\eqref{eq:20} implies that the RG-relevant mass term (set by $U_{2k_0}$) gaps out the pseudospin sector at a much higher energy scale than the inter-edge tunneling $t_\perp$. This justifies treating the $\vartheta_\tau$ field as pinned and focusing only on the RG flow of the two-body inter-edge tunneling in the effective low-energy theory (c.f. Eqs.~\eqref{eq:11} and \eqref{eq:12}), where single-particle inter-edge tunneling remains RG irrelevant. Practically, these weak inter-edge couplings could be realized by maintaining a large distance between the edges during the fabrication of the superlattice structure.

\section{Summary}
\par DQSHIs host DHESs, which yield unique $\pi$-SC and $\pi$-SDW phases driven by the long-range intra-edge forward scattering. These unconventional electronic orders are distinct from the conventional SC and SDW phases that arise from Fermi surface instabilities of Luttinger liquid physics. Inspired by the intricate phase competition in coupled quantum wire systems, we propose using the side surface of a double quantum spin Hall superlattice to realize an array of coupled DHESs by stacking double quantum spin Hall layers alternating with dielectrics. Our analysis demonstrates that such an array can provide a tunable platform for exploring competing 2D $\pi$-SC and $\pi$-SDW phases through the RG-relevant particle-particle and particle-hole inter-edge tunnelings. Further inter-edge tunnelings and higher-harmonic components of the HSLL parameter could be introduced by adjusting the inter-edge distance and dielectric constant through device fabrication to yield a new playground for investigating unconventional electronic orders more generally. 

\section*{Acknowledgement}
\par  The work at Northeastern University was supported by the Air Force Office of Scientific Research under award number FA9550-20-1-0322 and benefited from the resources of Northeastern University’s Advanced Scientific Computation Center, the Discovery Cluster, the Quantum Materials and Sensing Institute, and the Massachusetts Technology Collaborative award MTC-22032. C.-H.H. acknowledges the support from the National Science and Technology Council (NSTC), Taiwan, through Grant No.~NSTC-112-2112-M-001-025-MY3 and Academia Sinica (AS), Taiwan, through Grant No.~AS-iMATE-114-12.

\section*{Data Availability}
\par The results shown in Figs.~\ref{fig:04}, \ref{fig:06}, and \ref{fig:07} can be reproduced by the codes on GitHub \cite{github}, where the data that support the findings of this article are openly available.

\appendix
\section{Bosonization convention}\label{appx:A}
\par The DHESs host two left- and right-moving modes carrying different spin, forming two pairs of time-reversal partners in the time-reversal sector with the following bosonization expression \cite{PhysRevLett.121.196801}:
\begin{equation}\label{eq:S1}
\begin{split}
    R_{\mu,i,\uparrow}(r) & = \frac{\kappa_R}{\sqrt{2\pi a}}e^{i[\vartheta_{\mu,i}(r)-\varphi_{\mu,i}(r)]}, \\
    L_{\mu,i,\downarrow}(r) & = \frac{\kappa_L}{\sqrt{2\pi a}}e^{i[\vartheta_{\mu,i}(r)+\varphi_{\mu,i}(r)]},
\end{split}
\end{equation}
where $i$ is the edge index and $\mu=1,2$ indicates the time-reversal sector and $a$ is the cutoff. The bosonic fields satisfy the commutation relation $[\varphi_{\mu,i}(r),\vartheta_{\nu,j}(r')]=-i\frac{\pi}{2}\text{sgn}(r-r')\delta_{ij}\delta_{\mu\nu}$ for $\mu,\nu=1,2$ \cite{book:Giamarchi}, and can be further expressed in terms of charge ($c$) and pseudospin ($\tau$) sectors through:
\begin{align}\label{eq:S2}
    \begin{pmatrix}\varphi_{c,i}(r) \\ \varphi_{\tau,i}(r) \end{pmatrix} \equiv & \frac{1}{\sqrt{2}}\begin{pmatrix} 1 & 1 \\ 1 & -1 \end{pmatrix}\begin{pmatrix}\varphi_{1,i}(r) \\ \varphi_{2,i}(r) \end{pmatrix}
    \\ \begin{pmatrix}\vartheta_{c,i}(r) \\ \vartheta_{\tau,i}(r) \end{pmatrix} \equiv & \frac{1}{\sqrt{2}}\begin{pmatrix} 1 & 1 \\ 1 & -1 \end{pmatrix}\begin{pmatrix}\vartheta_{1,i}(r) \\ \vartheta_{2,i}(r) \end{pmatrix},
\end{align}
satisfying $[\vartheta_{\mu,j}(r'),\varphi_{\nu,i}(r)]=i\frac{\pi}{2}\text{sgn}(r-r')\delta_{ij}\delta_{\mu\nu}$ for $\mu,\nu=c,\tau$ \cite{PhysRevB.111.245145}.

\section{Second-order effect of the spin-orbit coupling (SOC)}\label{appx:A2}
\par In DHESs of DQSHIs, the edge Dirac points are solely protected by the spin-U(1) symmetry, so that any spin-U(1) symmetry-breaking perturbation, such as SOC, gaps the edge Dirac cones, see Fig.~\ref{fig:02}. Such a SOC term is described by the Hamiltonian:
\begin{equation}
    H_{\text{SOC}} = \lambda\int dr \bigg[ e^{i\theta}R_1^\dagger(r)L_2(r) + e^{-i(\theta+\pi)}R_2^\dagger(r)L_1(r) + \text{H.c.} \bigg],
\end{equation}
where $\lambda$ and $\theta$ describe the strength and phase of the SOC, respectively \cite{PhysRevB.110.035125}. This perturbation is RG relevant when the Fermi level is located at the edge Dirac points. However, if the Fermi level is above the edge Dirac point, which is the case considered here, it turns into a rapidly oscillating term and becomes irrelevant. Still, its second-order effect is relevant and has the following form:
\begin{equation}\label{eq:SOC_2nd_order}
    -\lambda^2 \int dr R_1^\dagger(r)L_2(r)L_1^\dagger(r)R_2(r) + \text{H.c.},
\end{equation}
which is proportional to the fermionic expression of the mass term in Eq.~\eqref{eq:01} of the main text. Equation \eqref{eq:SOC_2nd_order} contributes to this mass term via the second-order RG flow equation through
\begin{equation}
\frac{d\tilde{m}}{dl} = 2(1-K_\tau^{-1})\tilde{m} + \mathcal{O}(\lambda^2),
\end{equation}
where $\tilde{m}\equiv m/\hbar v$, $v$ is the fermi velocity, and $K_\tau$ is the LL parameter of the pseudospin sector \cite{PhysRevB.111.245145}.

\section{Helical sliding Luttinger liquid (HSLL)}\label{appx:B}
\par Here we derive the HSLL fixed point Hamiltonian and compute the parameters renormalized by the forward scatterings from the screened Coulomb interaction in the double quantum spin Hall superlattice in Fig.~\ref{fig:01} of the main text. 
\subsection{Screened Coulomb interaction in double quantum spin Hall superlattice}
\par To derive the screened Coulomb potential $V(\mathbf{x})$ in the double quantum spin Hall superlattice, we begin with the Poisson equation:
\begin{equation}\label{eq:B1}
    -\nabla^2V(\mathbf{x}) = \frac{1}{\epsilon}\rho(\mathbf{x}).
\end{equation}
For simplicity, we will assume similar values of the dielectric constant ($\epsilon$) for both the DQSHI and the dielectric material. Since the helical edge states are confined to a 2D plane, we place the charge density on $z=0$ plane, away from a metallic gate positioned at $z=D$. We thus rewrite Eq.~\eqref{eq:B1} as:
\begin{equation}\label{eq:B2}
    (q_x^2+q_y^2-\partial_z^2)V(\mathbf{q},z) = \frac{1}{\epsilon}\rho_{2D}(\mathbf{q})\delta(z).
\end{equation}
Here, we have Fourier transformed the density from $\rho(\mathbf{x})$ to $\rho_{2D}(\mathbf{q})\delta(z)$ due to the charge localization on the $z=0$ plane. By assuming the ansatz for the potential to be $V(\mathbf{q},z)=V_{2D}(\mathbf{q})V_\perp(z)$, we obtain:
\begin{equation}
    V_\perp(z) = A^{>(<)}e^{qz} + B^{>(<)}e^{-qz} \text{, for }z>0\,(z<0),
\end{equation}
where $q\equiv|\mathbf{q}|$. The metallic gate at $z=D$ imposes the following Dirichlet boundary conditions:
\begin{align}
    V_\perp(z=D) = V_\perp(z\to\infty) = 0
\end{align}
leading to $B^{<}=0$ and $B^{>}=-A^{>}e^{2qD}$. Since the potential is continuous across the $z=0$ plane, we also have $A^{<}=A^{>}(1-e^{2qD})$. To derive $A^{>}$, we apply Gauss's law on a flat surface enclosing $z=0$ plane, which yields $-\partial_zV_\perp\big|_{0^-}^{0^+}=\frac{1}{\epsilon}\frac{\rho_{2D}(\mathbf{q})}{V_{2D}(\mathbf{q})}$ and, hence,
\begin{equation}
    A^{>} = \frac{-e^{-2qD}\rho_{2D}(\mathbf{q})}{2q\epsilon V_{2D}(\mathbf{q})}.
\end{equation}
Using the above results, the potential on the $z=0$ plane is
\begin{equation}
    V(\mathbf{x},z=0) = \int \frac{d^2q}{(2\pi)^2} e^{i\mathbf{q}\cdot\mathbf{x}}\rho_{2D}(\mathbf{q})\frac{(1-e^{-2qD})}{2\epsilon q}.
\end{equation}

\subsection{HSLL Hamiltonian}
\par In DHESs, intra-edge potentials with non-zero momentum can renormalize the kinetic terms \cite{PhysRevB.111.245145}. However, since the zero-momentum component dominates the screened Coulomb potential, we neglect the contribution from the finite-longitudinal-momentum components to the renormalization of the kinetic terms for simplicity, as they do not significantly change the physics. For the double quantum spin Hall superlattice, the charge density can be expressed as:
\begin{equation}
    \rho_{2D}(\mathbf{x}) = e\sum_{m=1}^{N_\perp}\rho_m(x)\delta(y-md),
\end{equation}
where $N_\perp$ is the number of edges in the superlattice, and $d$ is the inter-edge distance. Then, the electron-electron interaction is:
\begin{align}
    H_{ee} & = \frac{1}{2}\int d^2x \rho_{2D}(x)V_{2D}(\mathbf{x}) \notag
    \\ & = \frac{e^2}{4\epsilon} \int d^2x \sum_{m=1}^{N_\perp}\rho_m(x)\delta(y-md) \notag
    \\ & \quad \quad \times \int \frac{d^2q}{(2\pi)^2} e^{i\mathbf{q}\cdot\mathbf{x}}\frac{(1-e^{-2qD})}{q} \notag
    \\ & \quad \quad \times \int d^2x' \sum_{m'=1}^{N_\perp}\rho_{m'}(x')\delta(y'-md)e^{-i\mathbf{q}\cdot\mathbf{x'}}. \label{eq:Hee}
\end{align}
Since we focus on contributions from zero-longitudinal-momentum $q_x\to0$ for the HSLL parameters and the renormalized velocities, the Hamiltonian in Eq.~\eqref{eq:Hee} is approximately:
\begin{align}
    H_{ee} & \cong \frac{e^2}{4\epsilon d}\int dx'\sum_{m,m'=1}^{N_\perp}\rho_{m}(x')\rho_{m'}(x') \notag
    \\ & \quad \quad \times\frac{1}{2\pi N_\perp}\sum_{q_y}e^{iq_y(m-m')d}\frac{(1-e^{-2|q_y|D})}{|q_y|}. \label{eq:Hee_1}
\end{align}
Here, we have used the discrete sum for the Fourier transform in the perpendicular direction. By defining the dimensionless transverse momentum $k_\perp\equiv q_yd$ and considering the limit of $N_\perp\to\infty$, we can further bosonize Eq.~\eqref{eq:Hee_1} into:
\begin{equation}
    H_{ee} = \frac{e^2}{2\pi^2\epsilon}\int \frac{dk_\perp}{2\pi} \int dr |\partial_r\varphi_{k_\perp}(r)|^2 \frac{(1-e^{-2|k_\perp|\frac{D}{d}})}{|k_\perp|}.
\end{equation}
Together with the kinetic energy terms, we get 
\begin{align}
    H & = \int \frac{dk_\perp}{2\pi}\int dr \big[ \frac{\hbar v}{2\pi} + \frac{e^2 \big( 1-e^{-2|k_\perp|\frac{D}{d}} \big)}{2\pi^2\epsilon |k_\perp|} \big] |\partial_r\varphi_{k_\perp}|^2 \notag
    \\ & \quad \quad \quad \quad \quad \quad  + \frac{\hbar v}{2\pi}|\partial_r\vartheta_{k_\perp}|^2, \label{eq:H_rk} 
\end{align}
which leads to the HSLL parameter and the renormalized velocity:
\begin{align}
    K_{k_\perp} & = \sqrt{\frac{1}{1 + \frac{e^2 \big( 1-e^{-2|k_\perp|\frac{D}{d}} \big)}{\hbar v\pi\epsilon |k_\perp|}}} ,
    \\ u_{k_\perp} & = v\sqrt{1 + \frac{e^2 \big( 1-e^{-2|k_\perp|\frac{D}{d}} \big)}{\hbar v\pi\epsilon |k_\perp|}} .
\end{align}
We further consider a large inter-edge distance ($d\gg D$) and the long-wavelength limit $|k_\perp|\ll1$, leading to the HSLL parameter and the renormalized velocity corresponding to the Hamiltonian in Eq.~\eqref{eq:H_rk} to be:
\begin{align}
    K_{k_\perp} & \cong \sqrt{\frac{1}{1 + \frac{2e^2D}{\hbar v\pi\epsilon d} - \frac{4e^2D^2|k_\perp|}{\hbar v\pi\epsilon d^2}}} \notag
    \\ & \cong \sqrt{\frac{1}{1 + \frac{2e^2D}{\hbar v\pi\epsilon d}}} \bigg( 1 + \frac{2e^2D^2|k_\perp|}{\hbar v\pi\epsilon d^2 \big( 1 + \frac{2e^2D}{\hbar v\pi\epsilon d} \big)}\bigg) \notag
    \\ & \cong \sqrt{\frac{1}{1 + \frac{2e^2D}{\hbar v\pi\epsilon d}}} \bigg( 1 + \frac{2e^2D^2}{\hbar v\pi\epsilon d^2 }\sin(|k_\perp|)\bigg) ,
    \\ 
    u_{k_\perp} & \cong v\sqrt{1 + \frac{2e^2D}{\hbar v\pi\epsilon d}}\equiv u.
\end{align}
recovering Eqs.~\eqref{eq:K0} and \eqref{eq:K1} of the main text.

\section{Renormalization-group (RG) flow equations}\label{appx:C}
\par To derive the RG flow equations for the phase diagram, we first Fourier transform the two-body inter-edge tunnelings in Eqs.~\eqref{eq:H_int_SC} and \eqref{eq:H_int_SDW} in the transverse direction:
\begin{align}
H_{\text{SC}}^{(\text{inter})} & = -\sum_{i\neq j}\frac{J_{\text{SC},|i-j|}}{(\pi a)^2} \notag
\\ & \times\int dr e^{-i2\sqrt{2}\int \frac{dk_\perp}{2\pi} [\vartheta_{k_\perp}(r)\sin(\frac{(i-j)}{2} k_\perp)\sin(\frac{(i+j)}{2} k_\perp)]} ,
\\ H_{\text{SDW}}^{(\text{inter})} & = -\sum_{i\neq j}\frac{J_{\text{SDW},|i-j|}}{(\pi a)^2} \notag
\\ & \times\int dr e^{-i2\sqrt{2}\int \frac{dk_\perp}{2\pi} [ \varphi_{k_\perp}(r)\sin(\frac{(i-j)}{2} k_\perp)\sin(\frac{(i+j)}{2} k_\perp)]},
\end{align}
where we have expanded the fields by
\begin{align}
\vartheta_{c,i}(r) & = \int \frac{dk_\perp}{2\pi} \vartheta_{k_\perp}(r)\cos(ik_\perp) ,
\\ \varphi_{c,i}(r) & = \int \frac{dk_\perp}{2\pi} \varphi_{k_\perp}(r)\cos(ik_\perp). 
\end{align}
Then, together with Eq.~\eqref{eq:HSLL}, the perturbative RG flow equations to the leading order contribution can be derived through a standard procedure \cite{PhysRevB.97.045415}:
\begin{align}
    \frac{d\tilde{J}_{\text{SC},N}}{dl} & = \big( 2-\eta_{\text{SC},N} \big)\tilde{J}_{\text{SC},N} , 
    \\ \frac{d\tilde{J}_{\text{SDW},N}}{dl} & = \big( 2-\eta_{\text{SDW},N} \big)\tilde{J}_{\text{SDW},N},
\end{align}
where $\tilde{J}_{...,N}\equiv J_{...,N}/\hbar u\pi^2$ are the dimensionless two-body inter-edge tunnelings and $N=|i-j|$ indicates the $N$th nearest neighbors and:
\begin{widetext}
\begin{align}
    \eta_{\text{SC},N} & = 2\sum_{I} \frac{1}{2}\int \frac{dk_\perp}{2\pi}\int \frac{dk_\perp'}{2\pi} \ln\bigg[ \frac{\langle \vartheta_{k_\perp}(r)\vartheta_{k_\perp'}(r) \rangle_0 L}{a} \bigg] 8\sin(\frac{I}{2} k_\perp)\sin(\frac{I}{2} k_\perp')\sin(\frac{N}{2} k_\perp)\sin(\frac{N}{2} k_\perp') \notag
    \\ & = 2\int \frac{dk_\perp}{2\pi}\int \frac{dk_\perp'}{2\pi} \ln\bigg[ \frac{\langle \vartheta_{k_\perp}(r)\vartheta_{k_\perp'}(r) \rangle_0 L}{a} \bigg] 4\pi\big(\delta(\frac{k_\perp}{2}-\frac{k_\perp'}{2}+2\pi m) - \delta(\frac{k_\perp}{2}+\frac{k_\perp'}{2}+2\pi m) \big)\sin(\frac{N}{2} k_\perp)\sin(\frac{N}{2} k_\perp') \notag
    \\ & = 2\int \frac{dk_\perp}{2\pi} \frac{1}{K_{k_\perp}}\sin^2(\frac{N}{2} k_\perp) \notag
    \\ & = \int \frac{dk_\perp}{2\pi} \frac{1-\cos(Nk_\perp)}{K_{k_\perp}}, 
    \\ \eta_{\text{SDW},N} & = 2\int \frac{dk_\perp}{2\pi}\int \frac{dk_\perp'}{2\pi} \ln\bigg[ \frac{\langle \varphi_{k_\perp}(r)\varphi_{k_\perp'}(r) \rangle_0 L}{a} \bigg] 4\pi\big(\delta(\frac{k_\perp}{2}-\frac{k_\perp'}{2}+2\pi m) - \delta(\frac{k_\perp}{2}+\frac{k_\perp'}{2}+2\pi m) \big)\sin(\frac{N}{2} k_\perp)\sin(\frac{N}{2} k_\perp') \notag
    \\ & = \int \frac{dk_\perp}{2\pi}K_{k_\perp} [1-\cos(Nk_\perp)].
\end{align}
\end{widetext}
For the nearest neighbors with $K_{k_\perp}=K_0[1 + K_1\sin(|k_\perp|)]$, we have 
\begin{align}
    \eta_{\text{SC},1} &  = \frac{1}{K_0}\frac{2\cos^{-1}(K_1)}{\pi\sqrt{1-K_1^2}} ,
    \\ \eta_{\text{SDW},1} & = K_{0}\big( 1 + \frac{2K_1}{\pi} \big).
\end{align}

\par The second-order effect of single-particle inter-edge tunneling $t_\perp$ can trigger two-body inter-edge tunnelings. Although single-particle tunneling is irrelevant, its second-order effects can introduce the relevant two-body inter-edge tunnelings \cite{book:Giamarchi}. The symmetry-allowed single-particle inter-edge tunneling has the form:
\begin{equation}\label{eq:S23}
    \sum_{i\neq j}\sum_{\mu=1,2}\int dr \,\, t_{\perp,|i-j|}\big[ R^\dagger_{\mu,i}(r)R_{\mu,j}(r) + L^\dagger_{\mu,i}(r)L_{\mu,j}(r) \big].
\end{equation}
Here, we consider a real inter-edge tunneling $t_{\perp,|i-j|}\in\mathbb{R}$ and ignore the fast-oscillating terms. Then, the second-order perturbation to the partition function includes the expectation value of:
\begin{widetext}
\begin{equation}\label{eq:S24}
    -\sum_{i\neq j}\sum_{\mu,\nu=1,2}\int dr \frac{t_{\perp,|i-j|}^2}{2}\big[ R^\dagger_{\mu,i}(r)R_{\mu,j}(r) + L^\dagger_{\mu,i}(r)L_{\mu,j}(r) \big]\big[ R^\dagger_{\nu,i}(r)R_{\nu,j}(r) + L^\dagger_{\nu,i}(r)L_{\nu,j}(r) + (i\leftrightarrow j)\big], 
\end{equation}
\end{widetext}
which considers only the second-order effects that do not introduce hoppings between the three edges. This can be further decomposed into two terms that contribute to the two-body inter-edge tunnelings:
\begin{widetext}
\begin{align}
    T_1 & = -\sum_{i\neq j}\sum_{\mu,\nu=1,2}\int dr  \frac{t_{\perp,|i-j|}^2}{2}\big[ R^\dagger_{\mu,i}(r)R_{\mu,j}(r)L^\dagger_{\nu,j}(r)L_{\nu,i}(r) + L^\dagger_{\mu,i}(r)L_{\mu,j}(r)R^\dagger_{\nu,j}(r)R_{\nu,i}(r) \big] , \label{eq:S25}
    \\ T_2 & = -\sum_{i\neq j}\sum_{\mu,\nu=1,2}\int dr \frac{t_{\perp,|i-j|}^2}{2}\big[ R^\dagger_{\mu,i}(r)R_{\mu,j}(r)L^\dagger_{\nu,i}(r)L_{\nu,j}(r) + L^\dagger_{\mu,i}(r)L_{\mu,j}(r)R^\dagger_{\nu,i}(r)R_{\nu,j}(r) \big]. \label{eq:S26}
\end{align}
\end{widetext}
The full two-body inter-edge tunneling is $T=T_1+T_2$, which can be decomposed into two parts upon bosonization:
\begin{widetext}
\begin{align}
    T_{\text{SC}} & = -\sum_{i\neq j}\sum_{\mu=1,2}\int dr  \,\, t_{\perp,|i-j|}^2 \big[ e^{-i2\vartheta_{\mu,i}}e^{i2\vartheta_{\mu,j}} + e^{-i2\varphi_{\mu,i}}e^{i2\varphi_{\mu,j}} \big] \notag
    \\ & \propto -\sum_{i\neq j}\int dr  \,\, t_{\perp,|i-j|}^2 \bigg\{ e^{-i\sqrt{2}(\vartheta_{c,i}-\vartheta_{c,j})}\cos[\sqrt{2}(\vartheta_{\tau,i}-\vartheta_{\tau,j})] + e^{-i\sqrt{2}(\varphi_{c,i}-\varphi_{c,j})}\cos[\sqrt{2}(\varphi_{\tau,i}-\varphi_{\tau,j})] \bigg\} \notag
    \\ & \to -\sum_{i\neq j}\int dr  \,\, t_{\perp,|i-j|}^2 e^{-i\sqrt{2}(\vartheta_{c,i}-\vartheta_{c,j})} \bigg\{ \cos[\sqrt{2}\vartheta_{\tau,i}]\cos[\sqrt{2}\vartheta_{\tau,j}] + \sin[\sqrt{2}\vartheta_{\tau,i}]\sin[\sqrt{2}\vartheta_{\tau,j}] \bigg\} , \label{eq:S27}
    \\ T_{\text{SDW}} & = -\sum_{i\neq j}\int dr  \,\, t_{\perp,|i-j|}^2 \bigg\{ e^{-i(\vartheta_{1,i}-\varphi_{1,i})}e^{i(\vartheta_{1,j}-\varphi_{1,j})} \big[ e^{-i(\vartheta_{2,i}+\varphi_{2,i})}e^{i(\vartheta_{2,j}+\varphi_{2,j})} + e^{-i(\vartheta_{2,j}+\varphi_{2,j})}e^{i(\vartheta_{2,i}+\varphi_{2,i})} \big]  + (1\leftrightarrow2)\bigg\} \notag
    \\ & = -\sum_{i\neq j}\int dr  \,\, t_{\perp,|i-j|}^2 \bigg\{ e^{-i\sqrt{2}(\vartheta_{c,i}-\vartheta_{c,j})}e^{i\sqrt{2}(\varphi_{\tau,i}-\varphi_{\tau,j})} +  e^{i\sqrt{2}(\varphi_{c,i}-\varphi_{c,j})}e^{-i\sqrt{2}(\vartheta_{\tau,i}-\vartheta_{\tau,j})} + (\varphi_\tau,\vartheta_\tau \leftrightarrow -\varphi_\tau,-\vartheta_\tau)\bigg\} \notag
    \\ & \to -\sum_{i\neq j}\int dr  \,\, t_{\perp,|i-j|}^2e^{i\sqrt{2}(\varphi_{c,i}-\varphi_{c,j})}\cos[\sqrt{2}(\vartheta_{\tau,i}-\vartheta_{\tau,j})] \notag
    \\ & = -\sum_{i\neq j}\int dr  \,\, t_{\perp,|i-j|}^2e^{i\sqrt{2}(\varphi_{c,i}-\varphi_{c,j})}\bigg\{ \cos[\sqrt{2}\vartheta_{\tau,i}]\cos[\sqrt{2}\vartheta_{\tau,j}] + \sin[\sqrt{2}\vartheta_{\tau,i}]\sin[\sqrt{2}\vartheta_{\tau,j}] \bigg\}, \label{eq:S28}
\end{align}
\end{widetext}
where we only keep the relevant terms upon pinning the $\vartheta_\tau$ field. 

According to Eqs.~\eqref{eq:S27} and \eqref{eq:S28}, $T_{\text{SC}}$ and $T_{\text{SDW}}$ contribute to the ($\pi$-)SC and ($\pi$-)SDW channels of the two-body inter-edge tunneling, respectively. These contributions are reflected in the RG flow equations, which incorporate terms of order $\mathcal{O}(t_\perp^2)$:
\begin{align}
    \frac{d\tilde{J}_{\text{SC},N}}{dl} & = \big( 2-\eta_{\text{SC},N} \big)\tilde{J}_{\text{SC},N} + \mathcal{O}(t_{\perp,N}^2) ,
    \\ \frac{d\tilde{J}_{\text{SDW},N}}{dl} & = \big( 2-\eta_{\text{SDW},N} \big)\tilde{J}_{\text{SDW},N} + \mathcal{O}(t_{\perp,N}^2).
\end{align}
As indicated by Eqs.~\eqref{eq:S27} and \eqref{eq:S28}, although $\tilde{J}_{\text{SC},N}(l=0)$ and $\tilde{J}_{\text{SDW},N}(l=0)$ vanish at the bare level, nonzero values for $\tilde{J}_{\text{SC},N}$ and $\tilde{J}_{\text{SDW},N}$ are generated during the RG flow due to finite $t_{\perp,N}(l=0)$ \cite{book:Giamarchi}. These RG equations thus provide leading-order estimates of $\tilde{J}_{\text{SC},N}(l\to0)$ and $\tilde{J}_{\text{SDW},N}(l\to0)$, which are used to calculate the characteristic length $L^*$ and temperature $T^*$, as discussed in the main text.

\bibliography{ref}
\end{document}